# LOCALIZED AND QUASI-LOCALIZED ENERGY LEVELS IN THE ELECTRON SPECTRUM OF GRAPHENE WITH ISOLATED SUBSTITUTIONAL IMPURITIES

S.B. Feodosyev, V.A. Sirenko, E.S. Syrkin, E.V. Manzhelii, I.S. Bondar,
*B.Verkin Institute for Low Temperature Physics and Engineering of the NAS of Ukraine,
47 Nauky Ave., Kharkiv, 61103, Ukraine*

K.A. Minakova,
*NTU "Kharkiv Polytechnic Institute", 2 Kyrpychova Str., Kharkiv, 61002, Ukraine*

Author to whom correspondence should be addressed: Sirenko@ilt.kharkov.ua

**Abstract**
Based on the calculation and analysis of local Green's functions of impurity atoms of low concentration in a two-dimensional graphene lattice, the conditions for the formation and characteristics of local discrete levels with energies lying outside the band of the quasi-continuous spectrum and quasi-localized states with energies near the Fermi one are determined. Specific calculations were performed for boron and nitrogen impurity atoms, which can actually replace carbon in graphite and graphene nanostructures. For a boron impurity that forms local discrete levels outside the band of the quasi-continuous spectrum, sufficiently simple analytical expressions for the conditions for their formation, energy, intensity at the impurity atom, and damping parameter are obtained. An analysis of the formation of states quasi-localized on nitrogen impurities with energy near the Fermi level in graphene nanostructures was carried out.

**Introduction**

It is well known that**,** in terms of its electron properties, graphene is a two-dimensional semiconductor with a zero band gap. This is due to the fact that its two-dimensional crystal lattice consists of two equivalent triangular sublattices that are "inserted" into each other in such a way that the vertices of one of them are at the centers of gravity of the other one, which in turn leads to a specific mutual arrangement of the electron spectral branches. The electron subbands formed by the symmetric and antisymmetric combination of wave functions on these two sublattices intersect at the edge of the Brillouin zone. Only along one of the directions in reciprocal space (direction $\Gamma K$, see Fig. 1*a*) spectral branches touch each other (in the point $K$). Along all other directions there is a gap of finite width between the branches. This manifests itself in the linear (relativistic) dependence of the electron dispersion when the quasi-wave vector $\boldsymbol{k}$ changes near the $K$ - point the first Brillouin zone (in graphene, $\varepsilon(K) = \varepsilon_F$ is the Fermi energy). The electron density of states (DOS) near the Fermi level has a characteristic V - shaped (Dirac) singularity (curve 1 in Fig. 1*b*), which corresponds to the cone-shaped energy spectrum near these points [1, 2].

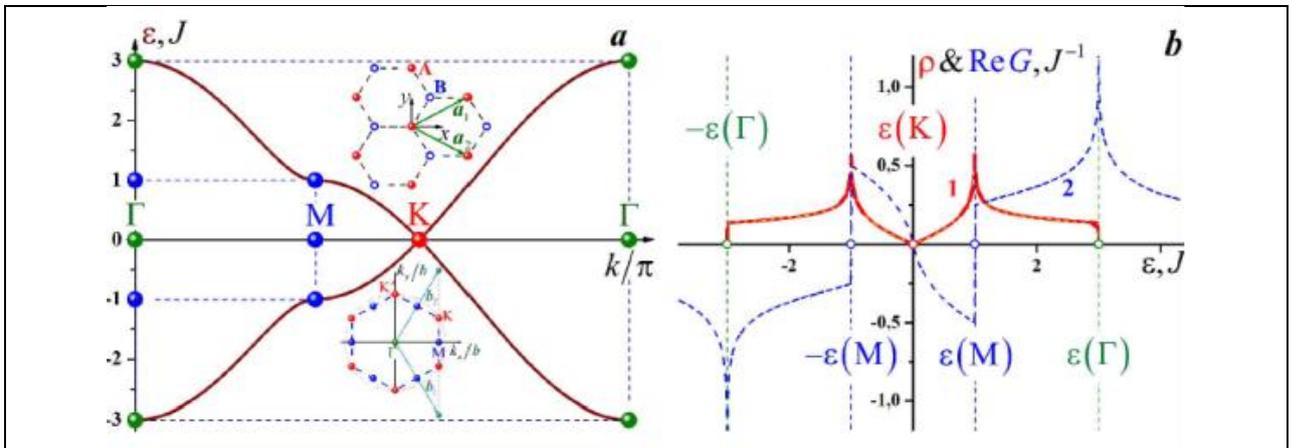

**Fig. 1:** (a) The dispersion curves of graphene along high symmetry directions: the unit cell of graphene and its first Brillouin zone with the positions of its high symmetry points are shown. (b) DOS of graphene (1) and the real part of its Green's function (2). Energy is measured from the Fermi level ε(K). On the fragment (a), the unit cell of graphene and its first Brillouin zone with the positions of its high symmetry points are shown.

Graphene monolayers cannot exist as planar formations in the free state, since in planar 2D crystals the rms amplitudes of atoms in the direction normal to the plane of the layer diverge even at T= 0 (see, for example, Ref. 3). It is possible to study and practically use graphene located on a certain substrate, which determines the stability of its flat shape (see, for example, Refs. 2, 4, and 5). Thus, when studying the electron properties of graphene, a dielectric (often silicon) substrate is used. This substrate does not change the electron spectrum of grafene. Only small flakes can be torn off from the substrate, which immediately acquire a corrugated shape [6] The presence of a substrate significantly expands the possibility of controlled introduction of various types of defects into graphene. For example, in graphene deposited on silicon, the formation of isolated vacancies is possible [7, 8] ]while in graphite, which is a set of graphene monolayers that weakly (van der Waals) interact with each other, vacancies are "healed", forming a stacking fault with a local axis of symmetry of the fifth order [9].The same thing applies to the incorporation of various kinds of impurities into graphene.

Impurity atoms embedded in graphene, at a certain ratio between the energies of particles in the sites and the interaction energy between atoms, can cause the appearance of impurity states outside the band of the quasi-continuous spectrum of the ideal structure. At low impurity concentrations, when the impurity can be considered as an isolated defect and its contribution to the total density of states is taken into account in the approximation linear in impurity concentration, these states can be represented as localized discrete levels (LDL). Such levels in various quasiparticle spectra have been known and have been studied for more than 70 years, but so far there has been no adequate description of them for sufficiently realistic (and even more so for real) models. The dependence of the conditions for the appearance and characteristics of LDL (the energy, the intensity on impurities at these energies, as well as the decay of the LDL intensity with distance from the impurity) on the parameters of the ideal lattice and the defect were determined only in the most general terms. At the same time, LDLs can serve as an important source of information about the defect structure and force interactions in real crystals. To extract it, it is useful to have analytical expressions relating the maincharacteristics of the LDL to the parameters of the defect and the main lattice. In the present work, such expressions are obtained for LDL caused by substitutional impurities in graphene (in particular, when carbon is replaced by boron).

In addition, it should be noted that the Dirac V-shaped singularity on the density of states of graphene determines the behavior near the Fermi level of the real part of the Green's function $\mathrm{Re}\, G(\varepsilon)$. $\mathrm{Re}\, G(\varepsilon)$ related to the density of states by the Kramers–Kronig relation [Fig. 1(b), curve 2]. Its shape indicates a high probability that, under the influence of various kinds of defects, elementaryexcitations will be localized near $\varepsilon_F$. The presence of such quasilocalized states increases the population of the Fermi level. This inspires a certain hope that by creating some defect structures in graphene nanoobjects, their transition to the superconducting state can be achieved. In this work, we consider the formation of quasilocalized states (QLS) in graphene near the Fermi level, which are caused by the replacement of carbon by nitrogen.

**Electron spectrum of graphene with defects**

The electron spectrum of graphene in the absence of strong magnetic fields can be described with a sufficiently high accuracy in the tight coupling approximation, and we can restrict ourselves to considering the interaction only between the nearest neighbors (see, for example, [4, 5, 10–12]). The corresponding Hamiltonian can be written as:

$$\hat{H} = \sum_i \varepsilon_i |i\rangle\langle i| - \sum_{i,j} J_{ij} |i\rangle\langle j|, \tag{1}$$

where indices $i$ and $j$ number the nodes of the two-dimensional lattice; $\varepsilon_i$ - **is** energy of a particle in a node $i$, and $J_{ij}$ is the so-called overlap integral of nodes and $i$ and $j$.

The electron density of states of graphene is presented on Fig. 2 by curve 1 (these calculations were performed using the method of Jacobian matrices [13-15]). In ideal graphene, atoms of different sublattices are physically equivalent. Each of the local Green's functions $G(\varepsilon,i) = \langle i|(\varepsilon\hat{I}-\hat{H})^{-1}|i\rangle$ coincides with the full Green's function $G(\varepsilon) = \lim_{N\to\infty}\frac{1}{N}\sum_{i=1}^{N}\langle i|(\varepsilon\hat{I}-\hat{H})^{-1}|i\rangle$. The Dirac V-shaped singularity present on the density of states at $\varepsilon = \varepsilon(K) = \varepsilon_F$ also determines the behavior of the real part of the Green's function $\operatorname{Re}G(\varepsilon)$ near the Fermi level. The Lifshitz equation [16], whose solutions are the energies of localized impurity levels both inside the band of the quasi-continuous spectrum (QCS) and outside this band (LDL), can be written as [3,13]:

$$\operatorname{Re}G(\varepsilon) = S(\varepsilon, \Lambda_{ik}). \qquad (2)$$

Where, function $S(\varepsilon,\Lambda_{ik})$ is determined by the perturbation operator $\hat{\Lambda}$ ($\Lambda_{ik}$ are matrix elements of this operator in the considered basis, the construction of which will be presented in the next section).. For a wide class of perturbations caused by defects, the shape of the curve $\operatorname{Re}G(\varepsilon)$ (curve 2 on Fig.1b) on the interval $[-\varepsilon(M),\varepsilon(M)]$ ensures the existence of solutions of Eq. (2) corresponding to the energies of quasi-localized states. In the considered model $\varepsilon(M) = J$.

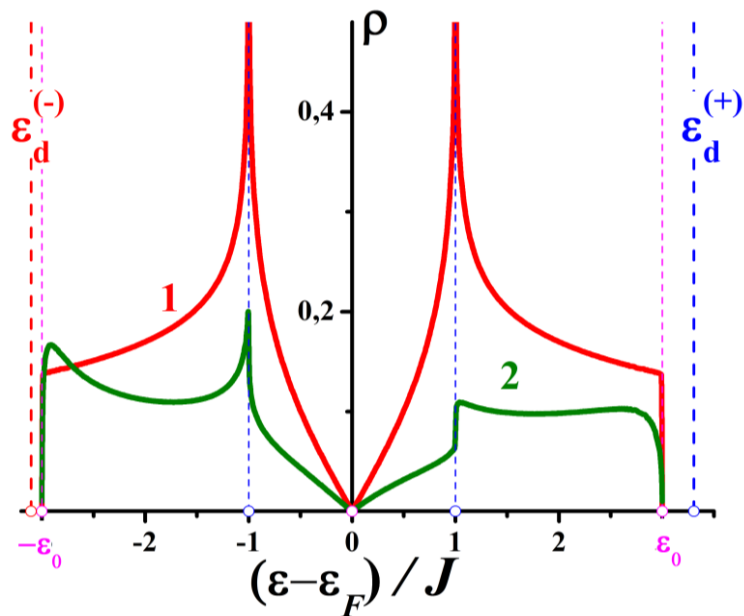

**Fig. 2:** The density of electron states of ideal graphene (curve 1) and the local density of states of the B atom, which is an isolated substitutional impurity in graphene (curve 2).

Local spectral densities $\rho(\varepsilon,i) \equiv \frac{1}{\pi}\lim_{\gamma\downarrow 0}\operatorname{Im}G(\varepsilon+i\gamma,i)$ of such impurity atoms were calculated in [5]. For an isolated substitutional impurity that differs from the main lattice atom by the energy values at the impurity site $i = 0$ ($\varepsilon_0 = \tilde{\varepsilon}$) and the overlap integral $J_{i0} = (1+\eta)J$ the function $S(\varepsilon, \tilde{\varepsilon}, \eta)$, calculated by the formula given in [13] has the form:

$$S(\varepsilon, \tilde{\varepsilon}, \eta) = \frac{(1+\eta)^2}{\tilde{\varepsilon} + \varepsilon\eta(2+\eta)}. \tag{3}$$

For boron impurity: $\eta \approx 0.5$ and $\tilde{\varepsilon} - \varepsilon(K) \approx 0.525J$ [5]. On Fig. 3, where the graphical solutions of equation (2) for a given impurity atom are given, the corresponding dependence $S(\varepsilon)$ is represented by curve 3. It can be seen that on the interval $[-\varepsilon(M), \varepsilon(M)]$ equation (2) has no solutions, and there are no quasi-localized states on this interval [5, 17]. The local Green's function (LGF) of the boron impurity has two poles outside the band of the quasi-continuous spectrum, which determine the local discrete levels (LDL) localized on the impurity atom. These levels are solutions of equation (2). As clearly seen in Fig. 2, the area under curve 2 representing the local impurity spectral density is smaller than the area under curve 1 (density of states of ideal graphene) by the sum of LGF residues at these poles.

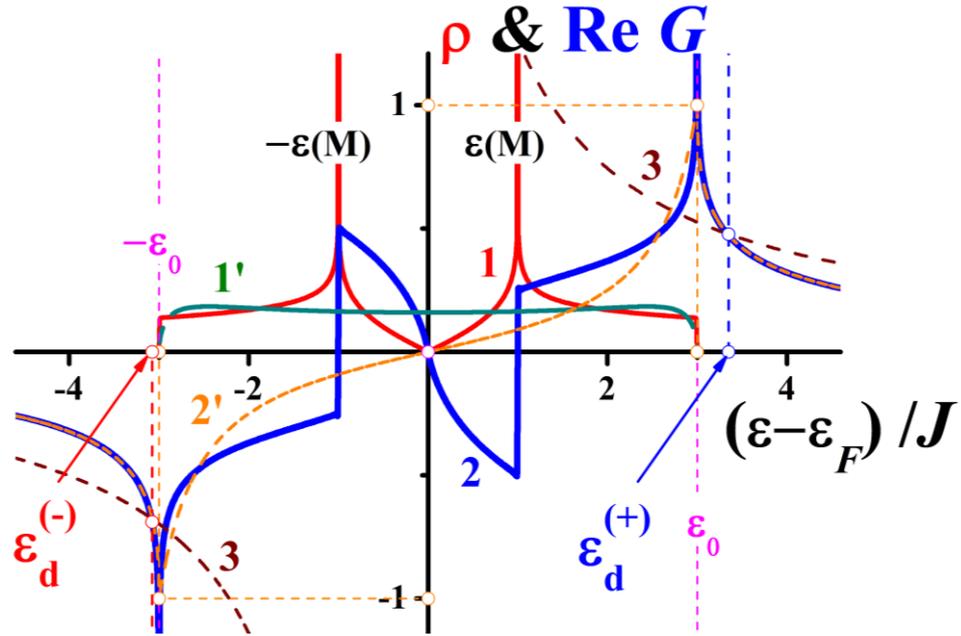

**Fig. 3:** Graphical solution of equation (2) for a boron substitution impurity in graphene: curve 1 is the density of electron states of ideal graphene, curve 2 is the corresponding real part of the Green's function. Curves 1′ and 2′ are the "two-time approximations" of these functions. Curve 3 is the function $S(\varepsilon)$, determined by the relation (3).

Local discrete levels are an important source of information about the defect structure and interatomic interactions (in particular, overlap integrals) in real structures. Its extraction can be significantly simplified in the presence of analytical expressions that relate the main characteristics of LDL (primarily their energies) with the parameters of the defect and the main lattice. In [18], such expressions were obtained for local vibrations in the phonon spectrum of a three-dimensional crystal. The analytical approximation of the main characteristics of local oscillations proposed in this paper was based on the rapid convergence in the basis of the Jacobian matrix method [13–15] of the real part of the Green's function outside the band of the quasi-continuous spectrum.

**Analytical approximation of the characteristics of local discrete levels in the electron spectrum of graphene**

First, let us briefly outline the basics of the method of Jacobian matrices to the extent necessary to understand the used classification of the eigenfunctions of the Hamiltonian (1). This method allows us to directly calculate the local partial Green's functions of the system under consideration, corresponding to some perturbation of one or several atoms, without finding the dispersion laws. This perturbation is given by the so-called generating vector $\vec{h}_0 \in H$. $H$ is the space of electron excitations of atoms of the given system. Its dimension is $qN$, where $N \to \infty$ is number of atoms in the system, and $q = 1, 2, 3$ is dimension of the displacement of an individual atom. Space vectors $H$ will be indicated by an arrow above the symbol, and "ordinary" $q$-dimensional vectors - as usual, in *bold italics*. In the space of scalar atomic excitations considered in this article $q = 1$.

If we construct the sequence $\{\hat{H}^n \vec{h}_0\}_{n=0}^{\infty}$, using the generating vector $\vec{h}_0 = \lambda^{-1} \sum_{j=1}^{\lambda} |j\rangle$ ($\lambda$ is the number of excited atoms) and the Hamiltonian (1), then the linear span spanned by its vectors forms in space $H$ some cyclic subspace invariant with respect to the operator $\hat{H}$. This subspace contains all atomic excitations generated by the vector $\vec{h}_0$. The corresponding local (partial) Green's function is defined as the matrix element $G_{00}(\lambda) \equiv \left(\vec{h}_0, \left[\varepsilon\hat{I} - \hat{H}\right]^{-1} \vec{h}_0\right)$, where $\varepsilon$ is the operator $\hat{H}$ eigenvalue. The value $\rho(\lambda) \equiv \pi^{-1} \operatorname{Im} G_{00}(\lambda)$ is called the spectral density generated by the initial excitation $\vec{h}_0$. In the basis $\{\vec{h}_n\}_{n=0}^{\infty}$, obtained by orthonormalization of the sequence $\{\hat{H}^n \vec{h}_0\}_{n=0}^{\infty}$, operator (1) is represented as a tridiagonal (Jacobian) matrix (or *J*-matrix). It has a simple spectrum, which greatly simplifies the determination of the partial Green's functions and spectral densities. As can be seen, this method does not explicitly use the translational symmetry of a crystal. It makes the J-matrix method exceptionally effective when considering systems with violated translational symmetry. The method of Jacobian matrices turns out to be especially effective when considering systems with a simply connected band of a quasi-continuous spectrum $D$. In this case, as the rank of the J-matrix increases ($n \to \infty$), its diagonal elements $a_n$ tend to the value $a$ corresponding to the middle of the band $D$, while the off-diagonal elements $b_n \to b$ (one quarter of the band width ). For the local Green's function (LGF) corresponding to the excitations of one or more atoms, which are given by the generating vector $\vec{h}_0$, in the J-matrices method the following expressions are obtained:

$$G(\varepsilon, \vec{h}_0) = \left(\vec{h}_0, \left[\varepsilon \hat{I} - \hat{H}\right]^{-1} \vec{h}_0\right) = \lim_{n \to \infty} \frac{Q_n(\varepsilon) - b_{n-1} Q_{n-1}(\varepsilon) K_\infty(\varepsilon)}{P_n(\varepsilon) - b_{n-1} P_{n-1}(\varepsilon) K_\infty(\varepsilon)}, \quad (4)$$

where, $\hat{I}$ is the unit operator; polynomials $P_n(\varepsilon)$ and $Q_n(\varepsilon)$ are determined by the following recursive relations:

$$b_n \{P, Q\}_{n+1}(\varepsilon) = (\varepsilon - a_n)\{P, Q\}_n(\varepsilon) - b_{n-1}\{P, Q\}_{n-1}(\varepsilon) \quad (5)$$

under initial conditions $P_{-1}(\varepsilon) = Q_0(\varepsilon) = 0$; $P_0(\varepsilon) \equiv 1$; $Q_1(\varepsilon) \equiv b_0^{-1}$; function $K_\infty(\varepsilon)$ corresponds to LGF of the operator, whose J- matrices elements are all equal to their limit values $a$ and $b$

$$K_\infty(\varepsilon) = 2b^{-2}\left[\varepsilon - a - Z(\varepsilon) \cdot \sqrt{(\varepsilon - a + 2b)(\varepsilon - a - 2b)}\right]; \quad (6)$$

$$Z(\varepsilon) \equiv \Theta(a - 2b - \varepsilon) + i\Theta(\varepsilon - a + 2b) \cdot \Theta(a + 2b - \varepsilon) - \Theta(\varepsilon - a - 2b). \quad (7)$$

The method of Jacobian matrices makes it possible to consider as regular degenerate perturbations a much larger number of perturbations of the phonon spectrum caused by various kinds of defects in the crystal than traditional methods (representation in the form of a superposition of plane waves, etc.) [13, 14]. In addition, perturbations that do not change the bandwidth of the quasi-continuous spectrum, and hence the asymptotic values of the elements of the J-matrix, can be considered as asymptotically degenerate regular perturbations [18]. These types of perturbations cover almost all perturbations of the phonon spectrum due to local defects. The calculation of the vibrational characteristics of such systems is carried out using the J-matrix method with the same accuracy as for the original ideal system.

In practice, it is usually possible to calculate J- matrix of Hamiltonian $\hat{H}$ with some finite rank $n$.

$$G(\varepsilon, \vec{h}_0) \approx \frac{Q_n(\varepsilon) - b_{n-1} Q_{n-1}(\varepsilon) K_\infty(\varepsilon)}{P_n(\varepsilon) - b_{n-1} P_{n-1}(\varepsilon) K_\infty(\varepsilon)} \quad (8)$$

Expression (8) is called the analytic approximation of the LFG, that is, its approximation by an analytic function. All dependencies in Fig. 1, as well as curves 1 and 2 in Fig. 2 are calculated by formula (8) using the J-matrix of rank $n = 600$ of the Hamiltonian (1), which corresponds to the calculation of these quantities by 1200 of their first moments. In the case we are considering, all the diagonal elements of the Jacobian matrix are equal to zero if we count the energy from the Fermi level ($\forall a_n = a = 0; b = \varepsilon_0/2$, where $\varepsilon_0 = 3J$ is the band half-width of the quasi-continuous spectrum). The degree of accuracy of the approximations presented in these figures can be estimated from the accuracy of their description of the van Hove singularities. Near these singularities, expression (8) converges to the true values of the real and imaginary parts of the LGF rather slowly.

The same local density of states and the real part of the LGF calculated by formula (8) at $n = 1$ are shown in Fig. 3 by curves $1'$ and $2'$, respectively. As can be seen in the band of the quasi-continuous spectrum, these dependences have very little in common with curves 1 and 2. Since, on the curve $1'$ there is no hint of $V$-shaped "Dirac" singularity at $\varepsilon = \varepsilon(K) = \varepsilon_F$, and on the curve $2'$ there are no nonmonotonicity (and nonanalyticity) on the interval $\left[-\varepsilon(M), \varepsilon(M)\right]$ and logarithmic singularities at the edges of the band of the quasi-continuous spectrum, that are inherent in $2D$ systems. However, outside the band of the quasi-continuous spectrum (including in the region of intersection of the real part of the LGF with curve (3)) curves 2 and $2'$ almost merge. If we substitute the LGF approximation (8) at $n = 1$ instead of the LGF into the Lifshitz equation (2), the resulting solutions give the values of the LDL energies with high accuracy. Moreover, these solutions can be easily found analytically. Approximation of LGF by formula (8) at $n = 1$ is called in [18] the two-moment approximation, since it corresponds to finding it from the first two moments. Indeed, from the condition of orthonormality of polynomials $P_n(\varepsilon)$ [13, 14] it follows:

$$\int_{-\varepsilon_0}^{\varepsilon_0} P_1(\varepsilon)\rho(\varepsilon)d\varepsilon = 0 \Rightarrow a_0 = \int_{-\varepsilon_0}^{\varepsilon_0} \varepsilon\rho(\varepsilon)d\varepsilon = M_1;$$

$$\int_{-\varepsilon_0}^{\varepsilon_0} P_1^2(\varepsilon)\rho(\varepsilon)d\varepsilon = \int_{-\varepsilon_0}^{\varepsilon_0} \frac{(\varepsilon^2 - a_0^2)}{b_0^2}\rho(\varepsilon)d\varepsilon = 1 \Rightarrow b_0 = \sqrt{M_2 - M_1^2}$$

It is more convenient to find the LDL without using Eq. (2), but to calculate them as LGF poles of the perturbed Hamiltonian $\tilde{G}(\varepsilon, \vec{h}_0) = \left(\vec{h}_0, \left[\varepsilon\hat{I} - \hat{H} - \hat{\Lambda}\right]^{-1} \vec{h}_0\right)$. In the two-moment approximation, for the subspace generated by excitation on an impurity atom (the index numbering the subspaces $\vec{h}_0$ further is omitted):

$$\tilde{G}(\varepsilon) = \frac{1}{\varepsilon - a_0 - b_0^2 K_\infty(\varepsilon)} \quad . \tag{9}$$

For the case of an isolated substitution impurity:

$$a_0 = \tilde{\varepsilon}; \quad b_0 = \sqrt{3}\tilde{J} = \sqrt{3}(1+\eta)J = \frac{1+\eta}{\sqrt{3}}\varepsilon_0; \tag{10}$$

from (10)

$$\tilde{G}(\varepsilon) = \frac{(1-\gamma)\varepsilon - a_0 + Z(\varepsilon)\gamma\sqrt{|\varepsilon^2 - \varepsilon_0^2|}}{R(\varepsilon)}, \tag{11}$$

let $R(\varepsilon) = (1-2\gamma)\varepsilon^2 - 2(1-\gamma)a_0\varepsilon + a_0^2 + \gamma^2\varepsilon_0^2$ and

$$\gamma \equiv b_0^2/2b^2 = 2(1+\eta)^2/3. \tag{12}$$

The local discrete energy levels are the poles of (11), that is, the roots of $R(\varepsilon)$

$$\varepsilon_d^{(\pm)} = \frac{(\gamma-1)a_0 \pm \gamma\sqrt{a_0^2 + (2\gamma-1)\varepsilon_0^2}}{2\gamma-1}. \tag{13}$$

Residues at these poles $\mu_0^{(\pm)} = \underset{\varepsilon=\varepsilon_d^{(\pm)}}{\mathrm{res}}\,\tilde{G}(\varepsilon)$ are called LDL intensities and determine the relative LDL "amplitude" on the impurity atom itself: $\mu_0^{(\pm)} = 1 - \pi^{-1}\int_{-\varepsilon_0}^{\varepsilon_0}\mathrm{Im}\,\tilde{G}(\varepsilon)d\varepsilon$. The non-zero intensity condition determines the LDL existence region. In the case under consideration:

$$\mu_0^{(+)} = \frac{\gamma a_0 + (\gamma-1)\sqrt{a_0^2 + (2\gamma-1)\varepsilon_0^2}}{(2\gamma-1)\sqrt{a_0^2 + (2\gamma-1)\varepsilon_0^2}} \cdot \Theta\left(\frac{a_0}{\varepsilon_0} - 1 + \gamma\right);$$

$$\mu_0^{(-)} = \frac{-\gamma a_0 + (\gamma-1)\sqrt{a_0^2 + (2\gamma-1)\varepsilon_0^2}}{(2\gamma-1)\sqrt{a_0^2 + (2\gamma-1)\varepsilon_0^2}} \cdot \Theta\left(\gamma - 1 - \frac{a_0}{\varepsilon_0}\right). \tag{14}$$

In [18], it was shown that

$$G_{mn}(\varepsilon, \vec{h}_0) = \left(\vec{h}_m, \left[\varepsilon\hat{I} - \hat{H}\right]^{-1}\vec{h}_n\right) = -P_m(\varepsilon)Q_n(\varepsilon) + P_m(\varepsilon)P_m(\varepsilon)G(\varepsilon, \vec{h}_0) \quad (m < n).$$

Hence it follows that the decay of LDL, i.e., the decrease in its intensity with increasing distance from the impurity atom (with increasing $n$), occurs as $\mu_n^{(\pm)} = P_n^2\left(\varepsilon_d^{(\pm)}\right) \cdot \mu_0^{(\pm)}$.

Using the method of mathematical induction, one can prove that

$$P_n\left(\varepsilon_d^{(\pm)}\right) = \sqrt{2\gamma} \cdot \left[\pm\frac{\sqrt{a_0^2 + (2\gamma-1)\varepsilon_0^2} \mp a_0}{\varepsilon_0(2\gamma-1)}\right]^n. \tag{15}$$

Intensites $\mu_n^{(\pm)}$ with increasing $n$ decrease by law $\mu_{n>0}^{(\pm)} = 2\gamma \cdot q^n \cdot \mu_0^{(\pm)}$ that is, starting from $n=1$, $\mu_n^{(\pm)}$ form an infinitely decreasing geometric progression, the denominator of which is

$$q^{(\pm)} = \left[ \frac{\sqrt{a_0^2 + (2\gamma-1)\varepsilon_0^2} \mp a_0}{\varepsilon_0(2\gamma-1)} \right]^2. \tag{16}$$

Summing these progressions, we see that $\sum_{n=0}^{\infty} \mu_n^{(+)} = \sum_{n=0}^{\infty} \mu_n^{(-)} = 1$, that is, the formation of each LDL is the formation of one quasi-particle outside the band of the quasi-continuous electron spectrum.

**Analysis of the obtained characteristics of local discrete levels. States localized on boron impurities**

Formulas (10), (12)-(14) and (16) give simple analytical expressions for the conditions for the existence of local discrete levels due to the presence of a substitutional impurity in graphene. On Fig. 4 on the plane $\{\tilde{\varepsilon}, \eta\}$ areas of existence of LDL are represented $\varepsilon_d^{(+)}$ - $\tilde{\varepsilon} > \varepsilon_0(1-\gamma) = \varepsilon_0 \left[ 1 - 2(1+\eta)^2/3 \right]$ and $\varepsilon_d^{(-)}$ - $\tilde{\varepsilon} < \varepsilon_0(\gamma-1) = \varepsilon_0 \left[ 2(1+\eta)^2/3 - 1 \right]$.

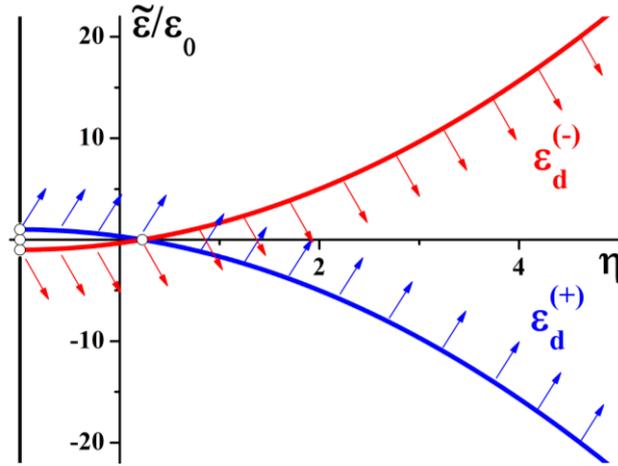

**Fig. 4:** Existence regions (shown by arrows) of discrete levels of substitutional impurities in graphene.

It can be seen that such levels exist in a very wide range of values $\tilde{\varepsilon}$ and $\eta$. The absence of LDL is possible only in a narrow range of values $\tilde{\varepsilon}$ at $\eta < \sqrt{3/2} - 1$. At $\eta > \sqrt{3/2} - 1$ there always exists at least one local discrete level. Realistically, lines delineating on a plane $\{\tilde{\varepsilon}, \eta\}$ region of existence of LDL in graphene must pass through the origin of coordinates, since the true dependence $\text{Re}\, G(\varepsilon) \to \infty$ at $\varepsilon \to \pm \varepsilon_0$. However, since this divergence is logarithmic, there is a certain threshold for any noticeable splitting of the LDL from the boundary of the quasi-continuous spectrum. Curves 2 and $2'$ on Fig. 2 coincide at once at $|\varepsilon| > \varepsilon_0$.

In Fig.5, the dependences of the LDL energies, intensities on impurities and damping parameters on the value $\eta$ for the boron impurity ($\tilde{\varepsilon} \approx 0.525 J$ [5]) are shown. The parameter $\eta$ characterizes the change in the overlap integral by an impurity atom (10). The solid lines show the characteristics of the LDL calculated using the approximation of the Green's function (9), that is, according to the analytical formulas (13), (14) and (16). Symbols ○ are results of numerical calculation of the dependencies $\varepsilon_d^{(\pm)}(\eta)$ and $\mu_0^{(\pm)}(\eta)$ with using the Green's function in the form (8) calculated from the Jacobian matrices of rank $n \geq 100$.. It can be seen that as the LDL energies tend to their threshold values $\varepsilon_d^{(\pm)} = \pm\varepsilon_0$, the intensities tend to zero, and the damping parameters tend to unity. Further increase in $\left|\varepsilon_d^{(\pm)}\right|$ is accompanied by an increase in $\mu_0^{(\pm)}$ and an increase in the level localization.

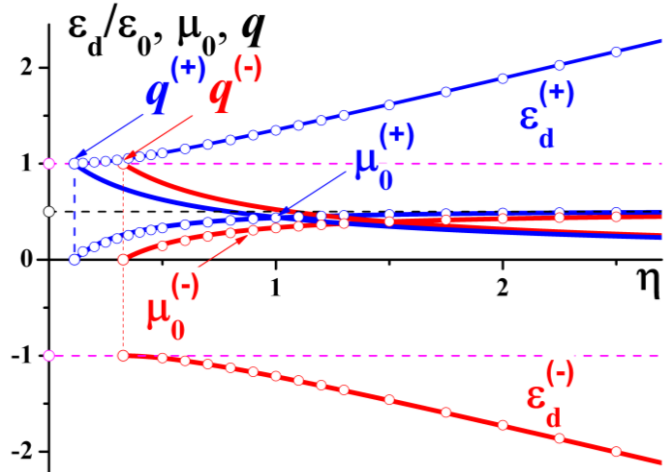

**Fig. 5:** Main LDL characteristics of substitutional impurities in graphene.

Good agreement between the results of numerical calculation of LDL characteristics using high-rank Jacobi matrices and their analytical description using Green's function approximation (9) makes it relatively easy to reconstruct defect parameters from known LDL characteristics. Relationships between defect parameters and LDL characteristics are given by formulas (13), (14). So experimental measurements (for example, by scanning tunneling microscopy) of values $\varepsilon_d^{(\pm)}$ makes it possible to determine the values of parameters $\tilde{\varepsilon}$ and $\eta$ using the formulas (10) (12) and (13). This is important for the synthesis of nanomaterials with predetermined spectral characteristics. As can be seen from formulas (12) and (14), with increasing $\eta$ intensity of LDL $\mu_0^{(+)} \to \mu_0^{(-)} \to 1/2$. That is, impurity levels cannot be completely localized on impurities, but also appear in the spectra of carbon atoms surrounding the impurities. This significantly increases the probability of experimental detection of such levels even at low impurity concentrations.

**Formation of quasi-local levels with energies close to the Fermi energy in the electron spectrum of graphene by small concentrations of impurities replacing carbon by nitrogen**

As noted in the Introduction, the behavior near the Fermi level of the real part of the Green's function $\operatorname{Re} G(\varepsilon)$ indicates a high probability that, under the influence of various kinds of defects, elementary excitations c will be localized near this level. The resulting increase in the population of the Fermi level may be sufficient for the transition of graphene nanoformations to the superconducting state.

For example in Fig 6a a graphical solution of the Lifshitz equation is given for the case when graphene contains an isolated impurity substitution of a nitrogen atom. The local spectral density of such an impurity atom was calculated in [5]. For an isolated substitutional impurity that differs from the main lattice atom by the energy values at the impurity site $i = 0$ ( $\varepsilon_0 = \tilde{\varepsilon} = \varepsilon_F - 0.525 J$ ) and the overlap integral $J_{i0} = 0.5 J$ function $S(\varepsilon, \tilde{\varepsilon}, \eta)$ (2) has the form shown in Fig. 6a (purple curves).

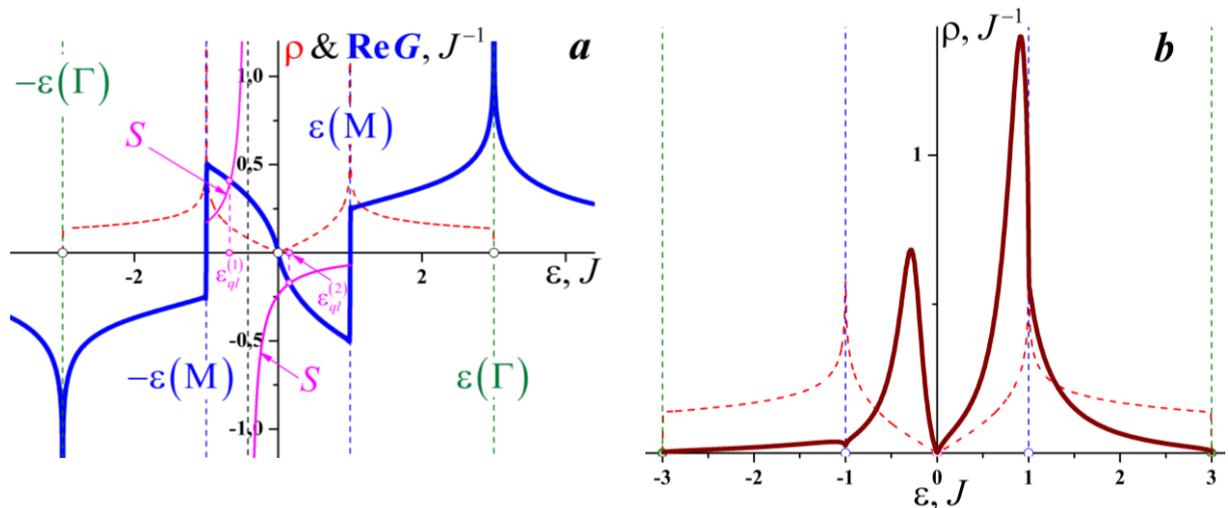

**Fig. 6:** The solution of equation (3) for the impurity substitution of carbon by nitrogen (fragment *a*) and LDOS of nitrogen atom in graphene (fragment *b*)

As can be seen from the figure 6, the equation (3) has solutions both on the interval $[-\varepsilon(M), \varepsilon(K)]$ that is point $\varepsilon_{ql}^{(1)}$, as well as in the interval $[\varepsilon(K), \varepsilon(M)]$ that is point $\varepsilon_{ql}^{(2)}$. Local densities of states (LDOS) of nitrogen impurities calculated by us by the method of Jacobian matrices (Fig. 6b) coincide with those calculated in [5]. They have quasi-local maxima in both these intervals. Due to the fact that the imaginary part of the Green's function differs from zero on these intervals, the locations of the quasi-local maxima differ from $\varepsilon_{ql}^{(1)}$ and $\varepsilon_{ql}^{(2)}$, however, the presence or absence of solutions to the Lifshitz equation on the interval $[-\varepsilon(M), \varepsilon(M)]$ for given defect parameters determines the presence or absence of quasi-localized states in this interval. Thus, for the case of boron impurity considered above, ($\tilde{\varepsilon}_0 \approx \varepsilon_F + 0.525 J$; $\eta \approx 0.5$) on the interval $[-\varepsilon(M), \varepsilon(M)]$ equation (3) has no solutions and there are no quasi-localized states on this interval. Note that the areas under the LDOS curves of the impurity nitrogen atom and the DOS of ideal graphene coincide, in contrast to the LDOS of boron impurity, which is less than the DOS of ideal graphene by the total LDL intensity.

**Conclusions**

Thus, in this work, we calculated and analyzed the local electron densities of states of impurity atoms of low concentration in a two-dimensional graphene lattice. The conditions for the formation of local discrete levels outside the band of the quasi-continuous spectrum and quasi-localized states near the Fermi energy are determined. For the conditions of occurrence and the main characteristics of local discrete levels, rather simple analytical expressions are obtained, which allow one to obtain information about the defect structure and interactions with defects from experimental data on measuring the characteristics of these levels. Note that the fact that two local discrete levels, separated from each other by more than the width of the quasi-continuous spectrum zone, are formed in boron-doped graphene, makes it possible to consider such a system as a "two-level" one. Such systems are of interest for nanoelectronics, in particular for quantum computing.

Analysis of the formation of states quasi-localized on nitrogen impurities near the Fermi level of grapheme was made. Increase in the population of the energy area near $\varepsilon_F$ under the influence of various kinds of defects [11, 19–22], as well as the enrichment of the frequency range with quasiparticles near the frequency $\omega(K)$ in spectrum of quasi-bending phonons [23–27] retains hope for the discovery of a superconducting transition in graphene nanostructures.

**Acknowledgment.**

Yuriy Alekseevich, in addition to the world-wide recognized high scientific authority, universal love and respect.

We wish the hero of the day good health, creative longevity, new outstanding scientific results and a peaceful sky above his head.

**References**

1. P.R. Wallace, *Phys. Rev*. **71**, 622 (1947).
2. K.S. Novoselov, A.K. Geim, S.V. Morozov, D. Jiang1, M.I. Katsnelson, I.V. Grigorieva, S.V. Dubonos, A.A. Firsov, *Nature* **438**, 197 (2005).
3. A.M. Kossevich, *The Crystal Lattice (Phonons, Solitons, Dislocations)*, Berlin: WILEY-VCH Verlag Berlin GmBH, 1999.
4. A.H. Castro-Neto, F. Guinea, N.M.R. Peres, K.S. Novoselov, and A.K .Geim, *Rev. Mod. Phys*. **81**, 109 (2009).
5. N.M.R. Peres, F.D. Klironomos, S.W. Tsai, J.M.D. Lopes dos Santos, A.H. Castro-Nero, Electrom "Waves in chemistry substituted graphene", *Europhys. Lett*. **80**, 67007 (2007).
6. J. C. Meyer, A. K. Geim, M. I. Katsnelson, K. S. Novoselov, T. J. Booth, and S. Roth, *Nature* **446**, 60 (2007).
7. Average Density of States in Disordered Graphene systems Shangduan Wu, Lei Jing, Qunxiang Li, Q.W. Shi, Le Chen, Xiaoping Wang, and Jinlong Ya, *Phys. Rev. B* **77**, 195411 (2008).
8. A. Feher, I.A. Gospodarev, V.I. Grishaev, K.V. Kravchenko, E.V. Manzhelii, E.S. Syrkin, S.B. Feodosyev, *Fiz. Nizk. Temp*. **35**, 862 (2009) [*Low Temp.Phys*. **35**, 655 (2009)].
9. L. Chen, Y. Zhang, Yui Shen, Self-healing in defective carbon nanotubes: a molecular dynamic study, *J. Phys. Cong. Matter* **9**, 38612-38616 (2007).
10. Yu.V. Skrypnyk and V.M. Loktev , *Phys.Rev.* **B73**, 241402 (2006).
11. Yu.V.Skrypnyk, V.M.Loktev, *Fiz. Nizk. Temp*. **42**, 863 (2016) [*Low Temp. Phys*. **42**, 679 (2016)].
12. C.Bena, S.A.Kivelson, *Phys. Rev.* **B 72**, 125432 (2005).
13. V.I. Peresada, A new computational method in the theory of harmonic vibrations of a crystal lattice, dissertation for the degree of doctor of physical and mathematical sciences (Kharkov, 1972).
14. V.I. Peresada, in: *Condensed Matter Physics*, FTINT AN UkrSSR, Kharkov (1968), p. 172.
15. V.I Peresada, V.N. Afanasyev, V.S. Borovikov, *Sov. Fiz. Nizk. Temp*. **1**, 461 (1975) [*Sov. J. Low Temp. Phys*. **1**, 227 (1975)].

16. I. M. Lifshits, *Reports of the USSR Academy of Sciences* **48**, 83 (1945).

17. A.Feher, E.Syrkin, S.Feodosyev, I.Gospodarev, K.Kravchenko, Quasi-particle spectra on substrate and embedded graphene monolayers, in: *"Physics and Application of Graphene - Theory"* (ed. S. Mikhailov*) "InTex" Open Access Publisher*, ISBN 978-953-307-996-7 Croatia, 2010, p. 94.

18. A. V. Kotlyar, and S. B. Feodosyev, *Fiz. Nizk. Temp*. **32**, 343 (2006) [*Low Temp. Phys*. **32**, 256 (2006)].

19. F. O. Ivashchyshyn, V. M. Maksymych, T. D. Krushelnytska, O. V. Rybak, B. O. Seredyuk, and N. K. Tovstyuk, *Fiz. Nizk. Temp*. **47**, 1165 (2021) [*Low Temp. Phys*. **47**, 1065 (2021)].

20. Yu.V.Skrypnyk, V.M.Loktev, *Fiz. Nizk. Temp*. **44**, 1237 (2018) [*Low Temp. Phys*. **44**, 1112 (2018)].

21. Yu.V.Skrypnyk, V.M.Loktev, *Fiz. Nizk. Temp*. **46**, 316 (2020) [*Low Temp. Phys*. **46,** 258 (2020)].

22. Yu.V.Skrypnyk, V.M.Loktev, *Fiz. Nizk. Temp*. **46**, 1028 (2020) [*Low Temp. Phys*. **46,** 869 (2020)].

23. I.A. Gospodarev, V.I. Grishaev, E.V. Manzhelii, E.S. Syrkin, S.B. Feodosyev, and K.A. Minakova, *Fiz. Nizk. Temp* **43**, 322 (2017) [*Low Temp. Phys*. **43**, 264 (2017)].

24. V. V. Eremenko, V. A. Sirenko, I. A. Gospodarev, E. S. Syrkin, S. B. Feodosyev, I. S. Bondar, A. Feher, and K. A. Minakova, *Fiz. Nizk. Temp*. **43**, 1657 (2017) [*Low Temp. Phys*. **43**, 1323 (2017)].

25. I.A. Gospodarev, V.A. Sirenko, E.S. Syrkin, S.B. Feodosyev, K.A. Minakova, *Fiz. Nizk. Temp*. **46**, 286 (2020) [*Low Temp. Phys*. **46**, 232 (2020)].

26. S. B. Feodosyev, I. A. Gospodarev, V. A. Sirenko, E. S. Syrkin, and S. Bondar and K. A. Minakova, *Fiz. Nizk. Temp*. 48, 137 (2022) [*Low Temp. Phys*. **48**, 121 (2022)].

27. S. B. Feodosyev, V. A. Sirenko, E. S. Syrkin, and I. A. Gospodarev, *Fiz. Nizk. Temp*. **48**, 710 (2022) [*Low Temp. Phys*. **48**, 628 (2022)].